\documentclass[a4paper,11pt]{article}
\usepackage{jheppub}
\usepackage[utf8]{inputenc}
\usepackage{amsfonts,amsmath,amssymb,amsthm}

\newcommand{\Cset}{\mathbb{C}}
\newcommand{\openone}{\mathbb{I}}

\newtheorem{lem}{Lemma}
\newtheorem{thm}{Theorem}
\newtheorem{defn}{Definition}

\begin{document}

\title{Formal Naive Dirac Operators and Graph Topology}
\author{G.M. von Hippel}
\emailAdd{hippel@uni-mainz.de}

\abstract{
Motivated by a recent conjecture of Misumi and Yumoto relating the number of
zero modes of lattice Dirac operators to the sum of the Betti numbers of the
underlying spacetime manifold, we study formal naive Dirac
operators on a class of graphs admitting such in terms of their zero modes. Our
main result is that for graphs on which translations commute, the conjecture of
Misumi and Yumoto can be shown and indeed can be strengthened to obtain bounds on
the individual Betti numbers rather than merely on their sum. Interpretations
of the zero modes in terms of graph quotients and of the representation theory
of abelian groups are given, and connections with a homology theory for such
graphs are highlighted.
}


\maketitle

\section{Motivation}

It has been known since the seminal work of Nielsen and Ninomiya
\cite{Nielsen:1980rz,Nielsen:1981xu,Nielsen:1981hk}
that the phenomenon of ``doublers'' for lattice formulations
of fermions is deeply related to topology. In the case of infinite regular
lattices, the crucial fact it that the Brillouin zone is a torus and thus has
Euler characteristic zero, and that for very general assumptions about the
fermion action the chiralities of the zero modes of the Dirac operator must add
to the Euler characteristic of the Brillouin zone, which implies that they must
come in pairs. This leaves open the question of how doublers on finite lattices
relate to the topology of those lattices, particularly if these lattices are
not just hypercubic grids, but more general graphs.

In a recent series of papers \cite{Yumoto:2023hnx,Yumoto:2023ums,Yumoto:2021fkm},
Misumi and Yumoto proposed the conjecture that
the number of doublers of a lattice Dirac operator is bounded from above by the
sum of the Betti numbers of the underlying spacetime manifold, with the bound
becoming sharp under some as yet not fully specified conditions. While the
claim is clearly empirically true for the torus, at least as long as
nearest-neighbor
derivatives are taken and all directions discretized with an even number of
points, it is less clear how this would work on a curved manifold, where the
curvature of the spin connection acts as an effective mass term and generally
prevents the appearance of zero modes altogether \cite{Flachi:2014jra}.

On the other hand, the sum of
the Betti numbers is not a very natural quantity (as opposed to their
alternating sum, which gives the Euler characteristic), and one is left to
wonder whether the doublers might not in fact encode more information about the
topology, up to and including the full set of Betti numbers. In this paper, I
set out a framework in which both the problem of curvature and even of a spin
connection is avoided altogether by considering formal Dirac operators on
graphs, while at
the same time giving access to the Betti numbers themselves, rather than merely
their sum, at least for a suitably constrained class of graphs.

The structure of this paper is as follows: in section \ref{sec:definitions} we
lay out our definitions and their consequences, in particular for the
eigenvalues of the doubler-count matrix which we define.
In section \ref{sec:graphint}, we interpret the
results so obtained in terms of connected components of a thinned version of
the graph, where the potential doubler modes can be understood as the vertices
of a quotient graph, with their mutual dependencies described by edges. In
section \ref{sec:groupint}, we specialize to the case of graphs admitting a
form of translational symmetry, which can be shown to be precisely the Cayley
graphs of finite abelian groups, whence we obtain a decomposition of such
graphs into box products of circulant graphs, which directly yields a count of
the number of doublers. Finally, in section \ref{sec:topint}, we show that for
the latter class of graphs a very simple homology theory can be constructed
that demonstrates them to be essentially tori and gives a derivation of a
strengthend form of the conjecture by Misumi and Yumoto.

\section{Definitions and Consequences}
\label{sec:definitions}

\subsection{Some graph-theoretic preliminaries}

We start out with defining our graph-theoretic setup.

A \emph{directed graph} $G$ is a pair $G=(V,E)$ of disjoint finite sets,
$V\cap E=\emptyset$, of vertices $v\in V$ and edges $e\in E$, together with a
pair of functions $\mathrm{src,snk}:E\to V$ assigning to each edge a vertex as
its source and its sink, respectively.
The \emph{in-degree} and \emph{out-degree} of a vertex $v\in V$ are the number
of edges $e\in E$ with $\mathrm{snk}(e)=v$ and $\mathrm{src}(e)=v$,
respectively.

We will demand that our graphs are \emph{simple}, i.e. 
$\mathrm{src}(e)\not=\mathrm{snk}(e)$ for all $e\in E$ (no loops), and that the map
$i:E\to V\times V$, $e\mapsto (\mathrm{src}(e),\mathrm{snk}(e))$ is injective
(no multiple edges).

Lastly, we will demand that our graphs are \emph{strictly directed}, i.e. for
$(v,v')\in i(E)$, $(v',v)\not\in i(E)$  (no bidirected edges).

We will call a simple strictly directed graph a \emph{digraph} for short.

For two directed graphs $G=(V,E)$, $G'=(V',E')$, we define their box product
$G\Box G'$ as having vertex set $V\times V'$ and edge set $V\times E'\cup
E\times V'$ with 
$\mathrm{src}(v,e')=(v,\mathrm{src}(e'))$,
$\mathrm{snk}(v,e')=(v,\mathrm{snk}(e'))$,
$\mathrm{src}(e,v')=(\mathrm{src}(e),v')$,
$\mathrm{snk}(e,v')=(\mathrm{snk}(e),v')$.

\subsection{Naive Dirac structures on graphs}

\begin{defn}
Let 
$G=(V,E)$ be a digraph.
We say that $G$ 
admits a naive Dirac structure in $d$ dimensions if
\begin{enumerate}
\item each vertex $v\in V$ has in-degree and out-degree both equal to $d$, and
\item there exists a $d$-colouring $\mu:E\to\{1,\ldots,d\}$ of the edges
such that each vertex $v\in V$ has for each $\nu\in\{1,\ldots,d\}$
exactly one incoming edge $e_{-\nu}(v)$ with $\mu(e)=\nu$ and one outgoing edge
$e_{+\nu}(v)$ with $\mu(e)=\nu$.
\end{enumerate}
We will call such a graph a \emph{$d$-Dirac graph} for short.
\end{defn}

Note that a $d$-Dirac graph is necessarily regular by the first condition,
which is required to ensure that the $d$-colouring works in such a way that an
analogue of the naive Dirac operator can be constructed.

We will need some additional constraints for some of our later results to hold,
and therefore we define

\begin{defn}
Let $G=(V,E)$ be a $d$-Dirac graph. We call a sequence $(e_1,\ldots,e_n)\in
E^n$, $e_k\not=e_l$ for $k\not=l$, a \emph{Dirac cycle}
if for all $k=1,\ldots n-1$ we have
$\mathrm{snk}(e_k)=\mathrm{src}(e_{k+1})$ and
$\mathrm{snk}(e_n)=\mathrm{src}(e_1)$, as well as
$\mu(e_k)=\nu$ for a fixed $\nu$ and all $k=1,\ldots,n$.
A Dirac cycle is even or odd depending on whether $n$ is even or odd.
\end{defn}

We note that Dirac cycles are essentially uniquely defined (up to a choice of
starting point) due to the constraint of having exactly one incoming and
outgoing edge of each colour at each vertex.

\begin{defn}
Let $G$ be a $d$-Dirac graph. We say that $G$ is
\begin{itemize}
\item \emph{fully even} if all Dirac cycles of $G$ are even,
\item \emph{commutative} if the maps $t_\mu:V\to V$,
$v\mapsto\mathrm{snk}(e_+\mu(v))$ commute, $t_\mu\circ t_\nu=t_\nu\circ t_\mu$.
\end{itemize}
\end{defn}

We first show that the box product of a $d$-Dirac graph and a $d'$-Dirac graph
is a $(d+d')$-Dirac graph:

\begin{lem} Let $G=(V,E)$ and $G'=(V',E')$ be digraphs admitting a naive Dirac
structure in $d$ and $d'$ dimensions, respectively. Then their box product
$G\Box G'$ admits a naive Dirac structure in $d+d'$ dimensions. If both $G$ and
$G'$ are commutative and/or fully even, then so is $G\Box G'$.

\begin{proof} From the definition, a vertex $(v,v')$ of $G\Box G'$ has $d'$ incoming
edges $(v,e_{-\nu}(v'))$, $d'$ outgoing edges $(v,e_{+\nu}(v'))$, 
$d$ incoming edges $(e_{-\nu}(v),v')$, and $d$ outgoing edges
$(e_{+\nu}(v),v')$, for a total of $d+d'$ incoming and outgoing edges each. The
map $\mu:V\times E'\cup E\times V'\to \{1,\ldots,d+d'\}$ with
$\mu(e,v')=\mu(e)$, $\mu(v,e')=d+\mu(e')$ then fulfils the requirements.

If both $G$ and $G'$ are commutative, for $\mu,\nu\le d$, we have $(t_\mu\circ
t_\nu)(v,v')=((t_\mu\circ t_\nu)(v),v')=((t_\nu\circ t_\mu)(v),v')=(t_\nu\circ
t_\mu)(v,v')$ and similarly for $\mu,\nu>d$, while for $\mu\le d<\nu$, we have
$(t_\mu\circ t_\nu)(v,v') = (t_\mu(v),t_\nu(v')) = (t_\nu\circ t_\mu)(v,v')$ as
the maps act only on $V$ and $V'$, respectively (and analogously for $\nu\le
d<\mu$). Since the Dirac cycles of $G\Box G'$ with colour $\mu\le d$ are copies
of those of $G$,  and the ones with colour $\mu>d$ are copies of those of $G'$,
$G\Box G'$ is fully even if both $G$ and $G'$ are.
\end{proof}
\end{lem}

\subsection{Derivative operators}

\begin{defn} Let $G=(V,E)$ be a digraph.
We consider the directed incidence matrix $\nabla_G\in\mathbb{R}^{E\times V}$ of $G$
given by
\begin{equation}
\left[\nabla_G\right]_{ev}=\left\{\begin{array}{rl}1,&v=\mathrm{snk}(e),\\-1,&v=\mathrm{src}(e),\\0,&\textrm{else,}\end{array}\right.
\end{equation}
where $e$ and $v$ range over $E$ and $V$, respectively.
\end{defn}

We note that $\nabla_G$ acts as a natural gradient operator on $G$, mapping
functions defined on the vertices to their gradients defined on the edges.
Considered as an operator acting on functions $f:V\to\Cset$, it acts as
\begin{equation}
(\nabla_Gf)(e) = f(\mathrm{snk}(e))-f(\mathrm{src}(e)).
\end{equation}

\begin{defn} Let
$G=(V,E)$ be a $d$-Dirac graph.
We introduce a set of formal variables $\gamma_\mu$, $\mu=1,\ldots,d$,
which take values in a ring $\mathbf{D}$. We will not care about representing
the $\gamma_\mu$ as matrices. The only property we
require for our purposes at first is that the set
$\{\gamma_\mu~|~\mu=1,\ldots,d\}$ is linearly independent. 
(Later on, we will require reality in the sense of $\gamma_\mu^\dag=\gamma_\mu$
and the canonical anticommutation relation
$\gamma_\mu\gamma_\nu+\gamma_\nu\gamma_\mu=2\delta_{\mu\nu}$ as well, so we can
take $\mathbf{D}$ to be the Clifford algebra on $\mathbb{R}^d$).

With these structures in place, we define the Dirac symmetrization matrix
$S_G\in\mathbf{D}^{V\times E}$ by
\begin{equation}
\left[S_G\right]_{ve}=\left\{\begin{array}{rl}\gamma_\mu,&e=e_{\pm\mu}(v)\\0,&\textrm{else}\end{array}\right.
\end{equation}
and use it to define the formal graph Dirac operator by
\begin{equation}
D_G = S_G\nabla_G.
\end{equation}
We moreover define the graph Laplacian by
\begin{equation}
\Delta_G = \nabla_G^\dag{}\nabla_G.
\end{equation}
\end{defn}

Considered as an operator acting on functions $f:V\to\Cset$, 
the graph Laplacian acts as
\begin{equation}
(\Delta_Gf)(v) = 2d f(v) - \sum_{\mu=1}^d \left[f(v_{+\mu})+f(v_{-\mu})\right],
\end{equation}
while the formal graph Dirac operator acts maps $f:V\to\Cset$ to
$(D_Gf):V\to\mathbf{D}$ with
\begin{equation}
(D_Gf)(v) = \sum_{\mu=1}^d \gamma_\mu\left[f(v_{+\mu})-f(v_{-\mu})\right] 
\end{equation}
where $v_{+\mu}=\mathrm{snk}(e_{+\mu}(v))$ and
$v_{-\mu}=\mathrm{src}(e_{-\mu}(v))$.
We note that the formal graph Dirac operator on a regular lattice in
$\mathbb{R}^d$ coincides with the naive Dirac operator when $\gamma_\mu$ are
taken to be the Dirac matrices and $f(v)$ is a taken to be a spinor rather than
a scalar.

In terms of the linear maps $T_\mu:\Cset^V\to \Cset^V$, 
$(T_\mu f)(v)=f(t_\mu(v))$, we have
\begin{equation}
D_G = \sum_{\mu=1}^d \gamma_\mu \left[T_\mu-T_\mu^\dag\right]
\end{equation}
and
\begin{equation}
\Delta_G = 2d\mathbb{I} - \sum_{\mu=1}^d \left[T_\mu+T_\mu^\dag\right].
\end{equation}

Note firstly that a constant is
always in the kernel of $\nabla_G$, since every row of $\nabla_G$ contains
exactly one $1$ and one $-1$, so that $\nabla_G$ always has a non-trivial
kernel, and hence so has $D_G$.

\begin{lem}
Let $G$ be a commutative $d$-Dirac graph. Then $D_G$ and $\Delta_G$ can be
simultaneously diagonalized.
\begin{proof}
Note that the commutator is
\begin{equation}
[D_G,\Delta_G] = \sum_{\mu,\nu=1}^d \gamma_\mu
[T_\mu-T_\mu^\dag,T_\nu+T_\nu^\dag]
=0
\end{equation}
since the identity matrix commutes with everything and $[T_\mu,T_\nu]=0$ from
the commutativity of $G$. Since $D_G$ and $\Delta_G$ commute, they can be
simultaneously diagonalized.
\end{proof}
\end{lem}

We note that this result would not hold in general without the assumption of
commutativity.

We will need some results about how the zero modes of $D_G$ and the eigenmodes
of $\Delta_G$ behave under box products, specifically:

\begin{lem}
Let $G=(V,E)$, $G'=(V',E')$ be graphs, and let $f:V\to\Cset$ and $g:V'\to\Cset$
satisfy $\Delta_Gf=\lambda f$ and $\Delta_{G'}g=\lambda'g$, respectively.
Define $f\otimes g:V\times V'\to\Cset$ by $(f\otimes g)(v,v')=f(v)g(v')$. Then
$\Delta_{G\Box G'}(f\otimes g) = (\lambda+\lambda')(f\otimes g)$.
\begin{proof} Noting that we can decompose $\Delta_{G\Box G'}$ into a sum 
$\Delta_G\otimes \openone+\openone\otimes\Delta_{G'}$,
we have $\Delta_{G\Box G'}(f\otimes g)=(\Delta_G f\otimes
g)+(f\otimes \Delta_{G'}g) = (\lambda+\lambda')(f\otimes g)$.
\end{proof}
\end{lem}

\begin{lem}
Let $G$ be a $d$-Dirac graph and let $G'$ be a $d'$-Dirac graph, and let $n$,
$n'$ be the dimensions of $\mathrm{ker}(D_G)$ and $\mathrm{ker}(D_{G'})$,
respectively. Then $\mathrm{dim~ker}(D_{G\Box G'})=nn'$.
\begin{proof} Acting on a function $(f\otimes g)$ with $f:V\to \Cset$ and
$g:V'\to\Cset$, $D_{G\Box G'}(f\otimes g)=(D_Gf\otimes g)+(f\otimes D_{G'}g)=0$
iff $D_Gf=0$ and $D_{G'}g=0$. Therefore, $\mathrm{ker}(D_{G\Box
G'})=\mathrm{ker}(D_G)\otimes \mathrm{ker}(D_{G'})$, and the claim follows.
\end{proof}
\end{lem}

\subsection{Doubler counts}

We now define the doubler-count matrix $W_G$, which tells us about how highly
oscillatory the zero modes of $D_G$ are:
\begin{defn}
Let
$G=(V,E)$ a $d$-Dirac graph.
Let $N_G$ be a matrix whose columns $n_i$ form an orthonormal basis of
$\ker(D_G)$, $(n_i,n_j)=\delta_{ij}$ with the natural scalar product
$(f,g)=\sum_{v\in V}f(v)^*g(v)$. We define the matrix
\begin{equation}
W_G = \frac{1}{4}N_G^\dag{}\Delta_GN_G.
\end{equation}
\end{defn}

For cyclic and circulant graphs, we can calculate the eigenvalues of $W_G$
directly:

\begin{lem}
Let $C_{2k}$, $k\ge 2$ be the cyclic graph on $2k$ vertices. Then $C_{2k}$ is a
$1$-Dirac graph, $\mathrm{dim~ker}(D_{C_{2k}})=2$, and $W_{C_{2k}}$ has
eigenvalues $0$ and $1$.
\begin{proof} Every vertex clearly has one incoming and one outgoing edge,
which can all be labelled by $\mu=1$. Pick a vertex $v_0$ and label the
vertices along the outgoing edges $v_1,\ldots v_{2k-1}$ in order;
a function $f:V\to\Cset$ with values $f(v_0)=f_0$ and $f(v_1)=f_1$ is in
$\mathrm{ker}(D_{C_{2k}})$ iff $f(v_{2i})=f_0$ and $f(v_{2i+1})=f_1$ for all
$i=1,\ldots,k-1$, whence $\mathrm{dim~ker}(D_{C_{2k}})=2$. An orthonormal basis
of $\mathrm{ker}(D_{C_{2k}})$ is provided by $f_0(v_i)=1/\sqrt{2k}$ and
$f_1(v_i)=(-1)^i/\sqrt{2k}$, where $\Delta_{C_{2k}}f_0=0$ and
$\Delta_{C_{2k}}f_1 = 4 f_1$, giving the stated eigenvalues for $W_{C_{2k}}$.
\end{proof}
\end{lem}

\begin{lem}
Let $C_{2k}(p)$ be the circulant graph $C_{2k}(p)$ on $2k$ vertices with jumps
$1=p_1<p_2<\ldots<p_d<k$ and $\{p_\mu,k\}$ coprime. Then $C_{2k}(p)$ is a
$d$-Dirac graph, $\mathrm{dim~ker}(D_{C_{2k}(p)})=2$, and $W_{C_{2k}(p)}$ has
eigenvalues $0$ and $d$.
\begin{proof} Every vertex clearly has $d$ incoming and outgoing edges, and we
can label the edges belonging to jump $p_\mu$ by $\mu$. Pick a vertex $v_0$ and
label the vertices along the Dirac cycle with label $\mu=1$ $v_1,\ldots
v_{2k-1}$ in order, assigning the vertex $v_i$ an even or odd parity according to
whether $i$ is even or odd. Then the $d$ Dirac cycles (one for each $\mu$) each
visit all vertices in alternating order of parity, whence a function $f:V\to\Cset$
with values $f(v_0)=f_0$ and $f(v_1)=f_1$ is in
$\mathrm{ker}(D_{C_{2k}(p)})$ iff $f(v_{2i})=f_0$ and $f(v_{2i+1})=f_1$ for all
$i=1,\ldots,k-1$, and $\mathrm{dim~ker}(D_{C_{2k}(p)})=2$.
An orthonormal basis of $\mathrm{ker}(D_{C_{2k}(p)})$ is provided by
$f_0(v_i)=1/\sqrt{2k}$ and $f_1(v_i)=(-1)^i/\sqrt{2k}$, where
$\Delta_{C_{2k}(p)}f_0=0$ and $\Delta_{C_{2k}(p)}f_1 = 4 d f_1$,
giving the stated eigenvalues for $W_{C_{2k}(p)}$.
\end{proof}
\end{lem}

We show some results regarding the eigenvalues of $W_G$:

\begin{lem}
Let $G$ be a $d$-Dirac graph. The eigenvalues of $W_G$ are contained in
$[0;d]$.
\begin{proof}
Note that the eigenvalues of $W_G$ are non-negative since $W_G=A^\dag{}A$ with
$A=\frac{1}{2}\nabla_GN_G$. Moreover, $W_G$ is a compression of $\Delta_G$, so
that its eigenvalues are bounded by those of $\Delta_G$ through
the Cauchy interlacing theorem. The eigenvalues of
$\Delta_G$ are known to be bounded by $\lambda_{\mathrm{max}}\le 4d$
from the Gershgorin disc theorem, since all Gershgorin discs are equal to
$B_{2d}(2d)$ since all diagonal entries are $2d$ and all row sums are zero.
Thus the eigenvalues of $W_G$ are less or equal to $d$.
\end{proof}
\end{lem}

Specializing to a commutative fully even $d$-Dirac graph, we can derive a
stronger result:

\begin{lem}
Let $G$ be a commutative fully even $d$-Dirac graph. Then the eigenvalues of
$W_G$ are integers.
\begin{proof}
First, we note that due to the linear independence of the $\gamma_\mu$,
$D_Gf=0$ implies that $f$ lies in the kernel of all of the operators
$T_\mu-T_\mu^\dag{}$. In particular, the restriction of $f$ to each Dirac cycle is
therefore equal to a zero mode of the one-dimensional Dirac operator on the
corresponding $C_{2k}$ subgraph and takes the same value on alternating points
along that Dirac cycle.
Secondly, note that the simultaneous diagonalizability of $D_G$ and $\Delta_G$
means that we can take the columns of $N_G$ to be eigenvectors of $\Delta_G$.
One column, in particular, which we can take w.l.o.g. to be the first,
can be taken to be the constant function $f(v)=1/\sqrt{|V|}$, and the other
columns in order to be orthogonal to the first must correspond to functions
summing to zero over all vertices.
We can then infer that these columns can be taken to correspond to functions
that take alternating values $(-1)^n/\sqrt{|V|}$ along the Dirac cycles of at
least one direction colour $\mu$.
For each such function, we fix an arbitrary vertex $v\in V$ with $f(v)=1$. Then
\begin{equation}
(\Delta_Gf)(v) = \sum_{\mu=1}^d \frac{2-2s_\mu(v)}{\sqrt{|V|}}
\end{equation}
where $s_\mu(v) = f(t_\mu(v))/f(v)\in\{\pm 1\}$, and hence
\begin{equation}
\frac{(\Delta_Gf)(v)}{f(v)} = \sum_{\mu=1}^d n_\mu(v)
\end{equation}
with $n_\mu(v)\in\{0,4\}$. Since $f$ is an eigenvector of $\Delta_G$, the
eigenvalues of $\Delta_G$ are thus integer multiples of $4$, and $W_G$ in this
basis is diagonal with integer entries. Since the eigenvalues of $W_G$ do not
depend on a particular choice of basis, they are therefore integers.
\end{proof}
\end{lem}

We can now show a Künneth-style formula for the $m_k$ of box products of
commutative fully even $d$-Dirac graphs:

\begin{thm}
Let $G$ be a $d$-Dirac graph and let $G'$ be a $d'$-Dirac graph, with both
being commutative and fully even, such that
$\mathrm{dim~ker}(D_G)=n$ and $\mathrm{dim~ker}(D_{G'})=n'$, and let the
multiplicity of the eigenvalue $k$ of $W_G$ and $W_{G'}$ be $m_k(G)$ and
$m_k(G')$, respectively. Then the multiplicity of the eigenvalue $k$ of
$W_{G\Box G'}$ is given by
\begin{equation}
m_k(G\Box G') = \sum_{\genfrac{}{}{0pt}{}{p,q}{p+q=k}} m_p(G)m_q(G').
\end{equation}
\begin{proof} 
Noting that $\mathrm{ker}(D_{G\Box G'})$ is spanned by tensor
products $(f\otimes g)$ with $D_Gf=0$ and $D_{G'}g=0$, and that $\Delta_G$ and
$\Delta_{G'}$ can be diagonalized in $\mathrm{ker}(D_G)$ and
$\mathrm{ker}(D_{G'})$, respectively, we can w.l.o.g. take our
basis $N_{G\Box G'}$ to consist of tensor products of eigenmodes of $\Delta_G$
and $\Delta_{G'}$. Since the eigenvalues of $\Delta_G$ and $\Delta_{G'}$ in
$\mathrm{ker}(D_G)$ and 
$\mathrm{ker}(D_{G'})$, respectively, are integers, the eigenvalues of
$\Delta_{G\Box G'}$ in $\mathrm{ker}(D_{G\Box G'})$ (and hence the eigenvalues
of $W_{G\Box G'}$) are then all pairwise sums of these integers. Therefore
$m_k(G\Box G')$ counts how many ways there are to form $k$ as a sum of
eigenvalues $p$ and $q$ of $W_G$ and $W_{G'}$, and the answer is that for any
decomposition of $k$ into $p+q$ there are $m_p(G)m_q(G')$ such ways, so that
the result follows.
\end{proof}
\end{thm}

One readily checks that for torus graphs $T_{2k}^d= C_{2k}^{\Box d}$ one
obtains $m_n(T_{2k}^d) = \genfrac{(}{)}{0pt}{}{d}{n}$ in agreement with the formula.

For an example of a $d$-Dirac graph that is not a box product,
the circulant graph $C_{2k}(p)$ with $1=p_1<p_2<\ldots<p_d<k$ and $\{p_\mu,k\}$
coprime has $m_0(C_{2k}(p))=m_d(C_{2k}(p))=1$ with all other $m_k(C_{2k}(p))=0$.

\section{Graph-theoretic interpretation}
\label{sec:graphint}

A natural interpretation of the Dirac operator is that of a ``square root'' of
the Laplacian. When we apply this reasoning to our case, we see that the formal
Dirac operator is clearly not the square root of the graph Laplacian. However,
we can define another Laplacian from squaring the formal Dirac operator as
\begin{equation}
\widehat{\Delta}_G = -D_G^\dag{} D_G,
\end{equation}
which can be decomposed as
\begin{equation}
\widehat{\Delta}_G = -\sum_{\mu,\nu=1}^d \gamma_\mu\gamma_\nu
\left[T_\mu-T_\mu^\dag{}\right]\left[T_\nu-T_\nu^\dag{}\right]
=\sum_{\mu=1}^d \left[2\mathbb{I}-T_\mu^2-(T_\mu^\dag{})^2\right]
\end{equation}
where we have now used
$\gamma_\mu\gamma_\nu+\gamma_\nu\gamma_\mu=2\delta_{\mu\nu}$ as well as
$T_\mu^\dag{}=T_\mu^{-1}$. We can thus identify $\widehat{\Delta}_G$ as the graph
Laplacian $\Delta_{G^\theta}$ of a ``thinned'' graph $G^\theta=(V,E')$ with
$i(E')=\{(v,t_\mu(t_\mu(v))) ~|~v\in V,\mu=1,\ldots,d\}$.

Any zero eigenvalue of $D_G$ is also a zero eigenvalue of
$\widehat{\Delta}_G$ and hence $\Delta_{G^\theta}$. On the other hand, the
meaning of zero eigenvalues of the graph Laplacian is well known: they
correspond to disconnected components of the graph. We therefore see that the
zero eigenvalues of $D_G$ correspond to disconnected components of $G^\theta$.

Since we can always take one zero mode to be constant on the whole graph, the
other zero modes can be taken to be functions that sum to zero over all vertices,
while being constant on each connected component. We can then without loss of
generality fix one component on which we take all zero modes to be positive,
i.e. we decompose $G^\theta=\bigoplus_{i=0}^{n-1}G_i$, where $G_i=(V_i,E_i)$,
and take $f_0(v)=1$ for all $v\in V$, and 
$f_k(v)=\pm 1$ for $v\in V$ with $f_k(v)=1$ for $v\in V_0$, $\sum_{v\in
V}f_k(v)=0$ for $k>0$.

Then $\frac{1}{4}f_k^\dag{}\Delta_G f_k$ counts the number of directions $\mu$ along
which the $G_i$ on which $f_k=-1$ are connected to those on which $f_k=+1$ when
embedding the $G_i$ back into $G$.

Identifying the vertices in each $G_i$ via an equivalence relation $\sim$, the
quotient graph $G^\delta=G/\sim$ then is the (undirected, since each edge is
bidirected, and weighted, since we count edges of each colour $\mu$ separately)
``doubler graph'' of $G$. Selecting a base point $v_0$ in $G^\delta$ and
identifying it with the constant mode, we can then identify the other vertices
as the doublers, with the corresponding eigenvalues of $W_G$ given by their
weighted distance from $v_0$.

\begin{defn}
We say that the Dirac cycles for colours $\mu$ and $\nu$ are
\emph{coupled} if $\mu$ and $\nu$ label the same edges of $G^\delta$, and
that $\mu$ \emph{dominates} $\nu$ if the Dirac cycles with colour $\nu$ become
loops in $G^\delta$ and $\mu$ labels edges going into the vertices to which
those loops are attached.
\end{defn}

It is apparent that if $\mu$ and $\nu$ are coupled, the Dirac zero modes will
either be oscillatory or constant along the Dirac cycles for both $\mu$ and
$\nu$, while if $\mu$ dominates $\nu$, Dirac zero modes can only oscillate
along the Dirac cycles of $\mu$ and must be constant along those of $\nu$.

\section{Representation-theoretic interpretation}
\label{sec:groupint}

There is a natural group-theoretic interpretation of commutative $d$-Dirac
graphs as Cayley graphs of abelian groups.

\begin{defn}
Let $\mathfrak{G}$ be a group, and $S$ a set of generators of $\mathfrak{G}$.
The \emph{Cayley graph} $\Gamma(\mathfrak{G},S)=(\mathfrak{G},E)$ is defined by
the existence of an edge labelled by $s_i\in S$ from $g$ to $gs_i$ for all
$g\in\mathfrak{G}$.
\end{defn}

Sabidussi's theorem \cite{Sabidussi:1958ams}
states that a digraph $G$ is the Cayley graph of some
group $\mathfrak{G}$ iff there is a simply transitive action of $\mathfrak{G}$
on the vertices of $G$ which is an automorphism of $G$.

A connected commutative $d$-Dirac graph $G$ carries a transitive action of
the abelian group generated by the translations $t_\mu$, since every vertex can
be reached from every vertex by a sequence of $t_\mu$ due to $G$ being
connected, and this action is simply transitive since
the abelian nature of the group means that the different paths
that could be taken all correspond to the same group element. The commutative
nature of the graph $G$ moreover means that this action is a graph
automorphism, since the existence of an edge with label $\nu$ between $v$ and
$v'$ implies $v'=t_\nu(v)$ and hence we have $t_\nu(t_\mu(v))=t_\mu(v')$ and
thus there is an edge with label $\nu$ between $t_\mu(v)$ and $t_\mu(v')$. 
We have therefore shown the

\begin{lem}A commutative $d$-Dirac graph $G$ is the Cayley graph of the abelian
group $\mathfrak{G}$ generated by the translations $t_\mu$.\qed\end{lem}

Noting that the fundamental theorem for finite abelian groups gives a decomposition
$\mathfrak{G}=\bigoplus_{k=1}^r \mathbb{Z}/q_k\mathbb{Z}$ into cyclic groups,
that the Cayley graph of a direct sum of abelian groups is the box product of
the Cayley graphs corresponding to the summands, and that
the Cayley graphs of cyclic groups are circulant graphs, we conclude from this
lemma the

\begin{thm} A commutative $d$-Dirac graph is the box product of circulant graphs.
\qed\end{thm}

Each circulant graph of even order yields at least one Dirac cycle, with other
Dirac cycles belonging to labels assigned to additional generators of the same
cyclic group either coupled or dominated, depending on whether the relative
exponent $m-n$ in the relation $t_\mu^m = t_\nu^n$ is even or odd.

This then leads naturally to an interpretation of Dirac zero modes on $G$ in
terms of the representation theory of abelian groups:
Given that the group $\mathfrak{G}$ generated by the translations of a
commutative $d$-Dirac graph is finite and abelian, all of its (complex)
irreducible representations (irreps for short) are one-dimensional by Schur's
first lemma, and there are therefore as many such complex irreps as
$\mathfrak{G}$ has elements (i.e. as $G$ has vertices).

We note that if $f$ living in an irrep $\rho$ is to be a Dirac zero mode,
$(T_\mu-T_\mu^\dag)f=0$ implies that $\rho(t_\mu)=\rho(t_\nu)^*$
and hence Dirac zero modes correspond to real representations $\rho$ of
$\mathfrak{G}$, which can only have $\pm 1$ as possible values of
$\rho(t_\mu)$. Indeed, every assignment of $\pm 1$ for $\rho(t_\mu)$ that is
consistent with any relations between $t_\mu$ and $t_\nu$ yields a real irrep of
$\mathfrak{G}$.

We also note that the labelling of the vertices of $G$ by the group elements of
$\mathfrak{G}$ implies that functions on vertices carry precisely the regular
representation of $\mathfrak{G}$, in which each irrep occurs exactly once for
the abelian case. We therefore obtain that each assignment of $\pm 1$ to any
$t_\mu$ that is not coupled to or dominated by another $t_\mu$ that has already
been assigned a value (and the assignments following from
the relations to the $t_\mu$ that are so coupled or dominated) yields a real
irrep of $\mathfrak{G}$ and hence a Dirac zero mode.

Specifically, let $r\le d$ be the number of maximal cyclic subgroups of
$\mathfrak{G}$. Then there are a total of $2^r$ Dirac zero modes, since we can
assign $\pm1$ independently to a generator of each.

\section{Relationship with homology of \texorpdfstring{$d$}{d}-Dirac graphs}
\label{sec:topint}

In the following, let $G=(V,E)$ always be a commutative, fully even $d$-Dirac
graph.

Since this structure is strong enough to allow us to equip $G$ with the structure of
a cubical complex, we can define a homology on a commutative, fully even
$d$-Dirac graph without requiring more powerful tools such as would be provided
by path homology \cite{GLMY:2020jms} or singular cubical homology of digraphs
\cite{Jimenez:2023dant}.
For $r<d$, we define the group $C_r(G)$ of $r$-chains on $G$
to be the abelian group generated by $\{e^{(r)}_{\mu_1\cdots\mu_r}(v) ~|~ v\in
V,\,1\le\mu_1<\ldots<\mu_r\le d\}$, where $e^{(0)}(v)=v$ and
$e^{(1)}_\mu(v)=e_{+\mu}(v)$, and we take $C_{-1}(G)=\{0\}$.

The boundary operators $\partial_r:C_r\to C_{r-1}$ are then the group
homomorphisms defined by
\begin{equation}
\partial_r e^{(r)}_{\mu_1\cdots\mu_r}(v) = \sum_{s=1}^r
(-1)^s\left[e^{(r-1)}_{\mu_1\cdots\widehat{\mu_s}\cdots\mu_r}(v)-e^{(r-1)}_{\mu_1\cdots\widehat{\mu_s}\cdots\mu_r}(t_{\mu_s}v)\right]
\end{equation}
where $\widehat{\mu_s}$ means that the index $\mu_s$ is to be omitted. One then
readily checks that the commutativity of $G$ implies that
$\partial_{r-1}\circ\partial_r=0$.

Considering the group homomorphism
$\kappa^{(r)}_{\mu_1\cdots\mu_r}:C_r\to\mathbb{Z}$ given by
\begin{equation}
\kappa^{(r)}_{\mu_1\cdots\mu_r}(e^{(r)}_{\mu_1\cdots\mu_r}(v))=1,
\end{equation}
we then see that any boundary $\partial_{r+1} c$, $c\in C_{r+1}$ fulfills
$\kappa^{(r)}_{\mu_1\cdots\mu_r}(\partial_{r+1} c)=0$. Therefore, a Dirac cycle
$c_\mu$ with colour $\mu$ and length $L$ satisfying $\kappa^{(1)}_\mu(c_\mu)=L$
cannot be a boundary. On the other hand, it is clearly a cycle in the sense of
$\partial_1 c_\mu=0$, implying that the first homology group $H_1(G)=\mathop{ker}
\partial_1/\mathop{im}\partial_2$ is not trivial, but has rank at least
$d$, implying $b_1(G)\ge d$.

Similarly, we can consider the sets $p_{\mu_1\mu_2}(v)$, $\mu_1<\mu_2$ of
vertices in $G$ obtained by fixing a base point $v\in V$ and considering all
walks starting from $v$ taking steps only along the $\mu_1$ and $\mu_2$
directions. These will consist of a union of Dirac cycles, and taking the sum
$c_{\mu_1\mu_2}(v)=\sum_{w\in p_{\mu_1\mu_2}(v)} e^{(2)}_{\mu_1\mu_2}(w)$ yields again a cycle
that is not a boundary, implying that $H_2(G)$ has rank at least $d(d-1)/2$.
Analogous constructions for any $r\le d$ yield $b_r(G)\ge \genfrac{(}{)}{0pt}{}{d}{r}$.

In fact, we know from the group-theoretic arguments above that all fully even commutative
$d$-Dirac graphs are tori (since circulant graphs are tori) and hence we have
in fact an equality $b_r(G)=\genfrac{(}{)}{0pt}{}{d}{r}$ instead of these lower bounds.

\begin{defn}
We say that the cycle $c_{\mu_1\cdots\mu_r}(v)$ \emph{witnesses} the oscillation of a
Dirac zero mode $f$ if $f(t_{\mu_i}(w))=-f(w)$ for all $w\in
p_{\mu_1\cdots\mu_r}(v)$ and all $i=1,\ldots,r$, and
that it is a \emph{maximal witness} if there is no superset
$\{\nu_1,\ldots,\nu_{r+1}\}\supset\{\mu_1,\ldots,\mu_r\}$ such that
$c_{\nu_1\cdots\nu_{r+1}}(v)$ also witnesses the oscillation of $f$.
\end{defn}

If $c_{\mu_1\cdots\mu_r}(v)$ is a maximal witness to the oscillation of $f$
with $\sum_{v\in V}f(v)=0$, we immediately have that $\Delta_Gf=4rf$.

We note that $c_{\mu_1\cdots\mu_r}(v)$ will fail to be a witness to the
oscillation of a Dirac zero mode if one of the $\mu_i$ is dominated by
another one. Similarly, a witness $c_{\mu_1\cdots\mu_r}(v)$ will fail to be a
maximal witness if there exists a $\nu\not\in\{\mu_1,\ldots,\mu_r\}$ such that
$\nu$ and one of the $\mu_i$ are coupled.

Since the witnesses are partially ordered by inclusion, we conclude that the maximal
number of distinct zero nodes is the number of potential maximal witnesses.
Since each such potential maximal witness is a cycle that is not a boundary,
their number is bounded by the sum of the dimensions of the homology groups,
i.e. of the Betti numbers, confirming the conjecture by Misumi and Yumoto.

Moreover, we have that for each $r\le d$, the number of potential maximal
witnesses is bounded by the number of inequivalent cycles
$c_{\mu_1\cdots\mu_r}(v)$, which is bounded by $b_r(G)$,
setting an upper bound on the number of Dirac zero modes
$f$ with $\Delta_Gf=4rf$ and hence on the multiplicity $m_r(G)$ of the
eigenvalue $r$ of $W_G$, so that we have the

\begin{thm}For a fully even commutative $d$-Dirac graph $G$, we have $m_r(G)\le
b_r(G)$.\qed\end{thm}

\section{Conclusions}

In this paper, we have shown that the conjecture by Misumi and Yumoto about Dirac
zero modes on graphs holds at least for the case of graphs admitting a naive
Dirac structure with commuting translations along the labelled directions,
which turn out to be box products of circulant graphs, and hence essentially tori.
We have moreover shown a strengthening of
their conjecture, namely that the eigenvalue multiplicities of the doubler count
matrix (which is essentially the Wilson term on the space of doublers) are
bounded by the individual Betti numbers of the corresponding torus. Due to the
restriction to commuting translations, this can be understood entirely in terms
of the structure and representation theory of finite abelian groups.

We note that the class of $d$-Dirac graphs includes the regular hypercubic
graphs $C_L^{\Box d-1}\Box C_T$ used in traditional lattice simulations, on
which the formal naive Dirac operator becomes the naive discretization of the
continuum Dirac operator, and the doubler matrix $W_G$ is (up to a
normalization factor) the Wilson term lifting the degeneracy of the doublers.

One interesting application of the results of the present paper could be to
consider circulant graphs $C_{2k}(p)$ on $2k$ vertices with jumps
$1=p_1<p_2<\ldots<p_d<k$ and $\{p_\mu,k\}$ coprime as a possibility for a novel
form of minimally doubled fermions \cite{Wilczek:1987kw,Borici:2008ym},
since they have exactly two Dirac zero modes.

To go beyond the present results, one would need to relax at least one of the
assumptions, and the most obvious of these is the requirement of commutativity,
without which most of the proofs fail.
It remains to be seen if there are any fully even \emph{noncommutative} $d$-Dirac
graphs that can serve as suitable proxies for the study of doublers on
e.g. spheres or higher-genus surfaces. The most obvious candidates are either
not fully even (such as the octahedral graph, any orientation of which with
two labels yields Dirac cycles of length three for one label,
or the Cayley graph \cite{Grochow:MOcomment}
of the alternating group $A_5$ with generators $(123)$ and $(12345)$) or
turn out to also be tori topologically (such as the Cayley graph of the
quaternion group $Q_8$). More research is needed to either identify and study
suitable classes of fully even noncommutative $d$-Dirac graphs,
or else to demonstrate their non-existence.

Another potential extension lies in generalizing the basic construction of the
formal naive Dirac operator to encompass not merely graphs with a $d$-colouring
of their edges, but some more general class of regular graphs that have natural
embeddings into $d$-dimensional manifolds (which in two dimensions would
include e.g. traingular and hexagonal grids). Finding the right generalization
is not straightforward, however, and further work is required in this
direction.

\acknowledgments
The author thanks Tatsuhiro Misumi for useful discussions and for an
attentive reading of an early draft version of this paper.

%\bibliographystyle{amsplain}
%\bibliography{betti}
\providecommand{\bysame}{\leavevmode\hbox to3em{\hrulefill}\thinspace}

\end{document}